\documentclass[12pt,a4pape]{article}
\usepackage{amsfonts}
\usepackage{amssymb}
\usepackage{amsmath,float}
\usepackage{graphicx}
\usepackage{color}
\usepackage{bbm}
\usepackage{physics}
		\newcommand{\deruno}[2]{\frac{d#1}{d#2}}

		\newcommand{\pdd}[1]{\partial_{#1}}
		
		\newcommand{\id}{\mathbbm{1}}

\begin{document}
\title{\bf One-dimensional scattering of fermions on $\delta$-impurities}
\author{J. Mateos Guilarte$^{1}$\footnote{guilarte@usal.es}, J. M. Munoz-Castaneda$^{2}$\footnote{jose.munoz.castaneda@uva.es} \\
 I.Pirozhenko$^{3,4}$\footnote{pirozhen@theor.jinr.ru} and L. Santamar\'\i  a-Sanz$^{2}$\footnote{lucia.santamaria@uva.es}\\
\small $^{1}$Departamento de F{\'\i}sica Fundamental and IUFFyM, University of Salamanca, Spain\\
\small $^{2}$Departamento de F{\'\i}sica Te{\'o}rica, At{\'o}mica y {\'O}ptica, Valladolid University, Spain\\
\small $^{3}$Bogoliubov Laboratory of Theoretical Physics, JINR, Dubna, Russia \\
\small $^{4}$Dubna State University, Dubna, Russia}

\date{\today}

\maketitle
\begin{abstract}
We study the spectrum of the 1D Dirac Hamiltonian encompassing the bound and scattering states of a fermion distorted  by a static background built from $\delta$-function potentials. After introducing the most general Dirac-$\delta$ potential for the Dirac equation we distinguish between \lq\lq  mass-spike\rq\rq  and \lq\lq electrostatic\rq\rq $\delta$-potentials. Differences in the spectra arising depending on the type of $\delta$-potential are studied in deep detail.
\end{abstract}\thispagestyle{empty}

\section{Introduction}
The Dirac equation with various relativistic potentials mimicking string-like or vortex-like backgrounds  has a long history. The best known example is the Aharonov-Bohm\cite{Aharonov1959} interaction of charged fermions with a field of an infinitesimally thin solenoid. The scattering of fermions on magnetic 'tHooft-Polyakov monopoles and on Abrikosov-Nielsen-Olesen  strings with consequent fractional fermion numbers and fermion number non-conservation was a hot topic in the late 70s and early 80s \cite{Jackiw1976,Rubakov1982}. The  cosmic strings predicted by grand unified theories  also appeared to interact with fermions (matter) via the Aharonov-Bohm mechanism~\cite{Wilczek1989}. The non-relativistic limit of  the scattering problem for spin-one-half particles in the Aharonov-Bohm potential in (1+2)  conical space was examined in \cite{Park-prd94,Andrade-prd2012}

It was  observed that in the case of a point magnetic vortex (Aharonov-Bohm interaction) one can either with gauge transformation reduce the problem to a Laplace equation with delta-potential or to a free Dirac equation with a special angular boundary condition~\cite{Jackiw95}. A radial  boundary condition  specifies the self-adjoint extension. From the operator theory  viewpoint the Dirac operator with relativistic point interaction ($\delta$-function potentials) and its self-adjoint extensions were considered in \cite{seba1989,Dabrowski1994}. In the last years the low-dimentional problems of  this kind  were investigated topologically with the Levinson theorem, which proved to be closely related to an index theorem \cite{Pankrashkin2014}.

A renewed interest to Dirac equation with  singular potentials  was  inspired by the appearance of the new 2D materials. 
Graphene in the field of a Aharonov-Bohm solenoid perpendicular to its plane  was considered in \cite{Jackiw2009}.  The induced current and induced charge density were calculated. Another example is a magnetic Kronig-Penney model for Dirac electrons in single-layer graphene developed in \cite{Ramezani_Masir_2009}. The model is a series of very high and very narrow magnetic $\delta$-function barriers alternating in sign.

In this paper we describe the distortion caused by impurities in the free propagation of fermionic fields in the $1+1$-dimensional Minkowski spacetime by means of Dirac $\delta$-point interactions. We elaborate on and develop further previous work on this subject in References \cite{bordag-jpa25,sundberg}. The matching conditions appropriated to define the $\delta$-potential inserted in a Dirac Hamiltonian restricted to a line were proposed some time ago in References \cite{lapidus,nogami}.

Our aim is to generalize the study carried out in References \cite{JMMCJMG, deltaprime, sine} to fermionic fields so that we can use the results in effective QFT models of 2D materials.

Throughout the paper we shall use natural units where $\hbar=c=1$ (henceforth, $L=T=M^{-1}$). In addition we will fix the Minkowski spacetime metric tensor to be: $\eta^{\mu\nu}\equiv{\rm diag}(+,-)$. Having done these choices, the Hamiltonian form of the Dirac equation, governing the dynamics of a free fermionic particle of mass $m$ moving on a line, reads
\begin{equation}
i\pdd{t}\psi(t,x)=\boldsymbol{H}_D^{(0)} \psi(t,x)= [-i\alpha \pdd{x} + \beta m] \psi(t,x), \label{fdireq}
\end{equation}
or, in covariant form:
\begin{equation*}
(i\gamma^\mu \pdd{\mu}- m\, ) \psi(x)=0.
\end{equation*}
For the fermionic anti-particle the dynamics is governed by the conjugate Dirac Hamiltoninan
\begin{equation}
i\pdd{t}\phi(t,x)=\overline{\boldsymbol{H}}_{ D}^{(0)}\phi(t,x)= [-i\alpha \pdd{x} - \beta m] \phi(t,x), \label{fdireq-ad}
\end{equation}
or, in covariant form
\begin{equation*}
(i\gamma^\mu \pdd{\mu}+ m\, ) \phi(x)=0.
\end{equation*}
Here, the $\{\gamma^\mu\}_{\mu=0,1}$ matrices close the Clifford algebra of ${\mathbb R}^{1,1}$ that can be minimally represented by \lq\lq pseudo-Hermitian\rq\rq two-by-two matrices: $\gamma^{\mu \, \dagger}=\gamma^0 \gamma^\mu \gamma^0$. Explicitly,
\begin{equation*}
 \{\gamma^\mu,\gamma^\nu\}\equiv\gamma^\mu\gamma^\nu+\gamma^\nu\gamma^\mu=2\eta^{\mu\nu} \, \, \, \Rightarrow \, \, \, \quad  \gamma^0\gamma^0=\id=-\gamma^1\gamma^1 \, \, , \, \, \, \gamma^0\gamma^1=-\gamma^1\gamma^0.
\end{equation*}
We shall need also the 1+1-dimensional analogue of the $\gamma^5$ matrix, denoted throughout this paper as $\gamma^2$:
\begin{equation*}
\gamma^2=\gamma^0\gamma^1.
\end{equation*}
The free Dirac Hamiltonians appearing in the formulas \eqref{fdireq}-\eqref{fdireq-ad} demand thus the definition of the $\alpha$ and $\beta$ Dirac matrices:
\begin{equation*}
\beta=\gamma^0, \qquad \qquad \alpha= \gamma^0 \gamma^1= \gamma^2.
\end{equation*}
In order to perform explicit calculations
we shall stick to the following choice of $\gamma$-matrices:
\begin{equation}\label{gammas}
\gamma^0=\sigma_3=\beta,\qquad \gamma^1=i\sigma_2, \qquad \gamma^2=\sigma_1 = \alpha,
\end{equation}
where $\sigma_1$, $\sigma_2$, $\sigma_3$ are the Pauli matrices:
\begin{equation*}
\sigma_1=\begin{pmatrix} 0 &  1 \\  1 & 0
\end{pmatrix},\quad
\sigma_2=\begin{pmatrix} 0 &  -i \\  i & 0
\end{pmatrix},\quad
\sigma_3=\begin{pmatrix} 1 &  0 \\  0 & -1
\end{pmatrix}.
\end{equation*}

Our choice of gamma matrices enables us to identify the three discrete transformations acting on the Dirac spinors:
\begin{enumerate}
\item Parity transformation ${\cal P}$: ${\cal P}\psi(x,t)=(-1)^p\gamma^0\psi(-x,t)$, where $p=0,1$, is the intrinsic parity of the particle
\item Time-reversal transformation ${\cal T}$: ${\cal T}\psi(x,t)=\gamma^0\psi^*(x,-t)$.
\item Charge conjugation transformation ${\cal C}$: ${\cal C}\psi(x,t)=\gamma^2\psi^*(x,t)$.
\end{enumerate}
Parity and time-reversal are symmetries of the free Dirac Hamiltoninan \eqref{fdireq} and its conjugate \eqref{fdireq-ad}. Charge conjugation, however, transforms the free Dirac Hamiltonian into its conjugate and viceversa. This property is the secret behind the common wisdom in Fermi field theory where negative energy fermions are traded by positive energy antifermions. More precisely: both $\boldsymbol{H}_D^{(0)}$ and $\overline{\boldsymbol{H}}_D^{(0)}$ have a Dirac sea of negative energy states. The idea is to form a complete set of spinors from the positive energy states of $\boldsymbol{H}_D^{(0)}$, fermions, and the positive energy eigen spinors of $\overline{\boldsymbol{H}}_D^{(0)}$, anti-fermions. In this spirit one checks that the solutions for the conjugate Dirac equation are related to the solutions of the Dirac equation by the charge conjugation transformation. Given the spectral problems
\begin{equation}
\boldsymbol{H}_D^{(0)} \psi_\omega(x)=\omega \psi_\omega(x), \quad \overline{\boldsymbol{H}}_D^{(0)} \phi_\omega(x)=\omega \phi_\omega(x),
\end{equation}
the eigenspinors are related through charge conjugation: $\phi_\omega=\gamma^2\psi^*_\omega(x)$. Equivalently, one finds that: ${\cal C} \boldsymbol{H}_D^{(0)}{\cal C}^{-1}= \overline{\boldsymbol{H}}_D^{(0)}$.

Consider next a bunch of relativistic Fermi particle propagating in (1+1)D Minkowski spacetime under the influence of a external time-independent classical background. The most general Dirac Hamiltonian describing this situation reads:
\begin{equation}
\boldsymbol{H}_D=\boldsymbol{H}_0+V(x)= -i  \alpha \partial_x + \beta m + V(x).
\end{equation}
The external potential comprises four types, see Ref. \cite{sundberg}:
\begin{equation}
V(x)= V_0(x)\mathbbm{1} + V_1(x) \alpha + V_2(x) \beta + V_3(x)\alpha \beta.
\end{equation}
In Reference \cite{sundberg} it is shown that:
\begin{itemize}
\item It is possible to assume $V_1(x)=0$ without loss of generality since it can be absorbed by a gauge transformation.
\item It is convenient to choose $V_3(x)=0$ to avoid interactions of the type $\bar{\psi} \gamma^2 V_3(x) \psi$ which are only consistent if $V_3$ is purely imaginary.
\end{itemize}
Hence, we shall focus our attention on background potentials of the form
\begin{equation}
V(x)= \xi(x)\,  \mathbbm{1} + M(x) \beta,
\end{equation}
leading to the following Dirac spectral problem
\begin{equation}
\boldsymbol{H}_D \psi(x)= \omega \psi(x)  \,\Rightarrow \,  [-i  \alpha  \partial_x+ \beta (m + M(x))]\psi(x)= [\omega-\xi(x)]\psi(x). \label{dirbacspec}
\end{equation}
In formula (\ref{dirbacspec}) the $\xi(x)$ potential clearly appears as an electrostatic potential, whereas the  potential energy $M(x)$ shows itself as a position dependent mass. Note that this last potential can be reinterpreted as an interaction of the Dirac field with a classical scalar field. Different elections for the electrostatic potential $\xi(x)$, and the mass-like potential $M(x)$ have been done in the last two decades: in Ref. \cite{akcay-pla09} it is studied the choice of $\xi(x)$ and $M(x)$ as Coulomb and quadratic {\it vetor potentials}, and in Refs. \cite{hiller-ajp02,nogami-ajp03}  the possibility of $\xi(x)$ and $M(x)$ being quadratic, linear, and other confining potentials  is considered.

We shall focus on external potentials localized in one point meaning that the propagating fermion finds an impurity at that point. Analytically we mimic the influence of the impurity on the fermion by a $\delta$-function potential. We thus choose:
\begin{equation}\label{gen-delta-v}
V(x)= \Gamma(q,\lambda) \delta(x);\quad \Gamma(q,\lambda)=q\mathbbm{1} + \lambda \beta.
\end{equation}
It is of note that the weight term $\Gamma(q,\lambda)$ multiplying the $\delta(x)$-function in formula (\ref{gen-delta-v}) is a $2\times 2$ matrix depending on two coupling constants: physically $q$ plays the role of a dimensionless electric charge{\footnote{We shall allow $q$ to vary as an angle proportional to the fine structure constant, which in a $1D$ space is $\alpha=\vert\frac{e^2}{m^2}\vert$, the electron charge times the particle Compton wavelength to the square.}} and $\lambda$ is also non dimensional, but plays the role of a scalar or gravitational coupling because it couples to the Fermi fields like a mass. Physically all this enables to interpret the most general form of the delta potential as a point charge plus a variable mass, in parallel to that taken in \cite{JMMCJMG} devoted to scalar field interactions with external $\delta$-plates.

Early distributional definitions of $\delta$-point interaction for Dirac fields were proposed in Refs. \cite{lapidus,nogami}. In these References the purely electrostatic fermionic Dirac-$\delta$ potential was defined through a matching condition of the form
\begin{equation}\label{elec-d}
\psi(0^+)=T_{E\delta}(q)\psi(0^-);\quad T_{E\delta}(q)=\mathbbm{1}\cos(q)-i \gamma^2\sin (q).
\end{equation}
Later, in Ref.\cite{sundberg} the matching condition \eqref{elec-d} was extended for the general $\delta$-potential \eqref{gen-delta-v}, following the approach of \cite{nogami}, to be:
\begin{eqnarray}\label{gen-delta-def}
&&\psi(0^+)=T_\delta(q,\lambda)\psi(0^-);\quad T_\delta(q,\lambda)=\exp\left(-i \gamma^2 \Gamma(q,\lambda) \right)\\
&&T_\delta(q,\lambda)= \mathbbm{1} \cos \Omega - \frac{i}{2}\sin \Omega \left[ \frac{\Omega}{q+\lambda}(\gamma^2+\gamma^1)+
\frac{q+\lambda}{\Omega}(\gamma^2-\gamma^1)\right],\nonumber
\end{eqnarray}
being $\Omega=\sqrt{q^2-\lambda^2}$.
It is straightforward to obtain the matrix that defines the mass-spike Dirac-$\delta$ potential:
\begin{equation}\label{eq9}
T_{M\delta}(\lambda)=T_\delta(0,\lambda)=\mathbbm{1}\cosh(\lambda)+i\gamma^1\sinh(\lambda).
\end{equation}
This last particular case is what is studied in detail in Ref.\cite{sundberg} regarding the Casimir effect induced by vacuum fermionic quantum fluctuations.

In order to comprehend the results of the calculations in the following sections it is convenient to take into account the transformation properties of the point-supported potential defined by equation \eqref{gen-delta-def} under parity (${\cal P}$), time-reversal (${\cal T}$), and charge conjugation (${\cal C}$).
\begin{itemize}
\item Taking into account that the the parity-transformed spinor $\psi^P\equiv{\cal P}\psi$ satisfies
\[
\psi^P(0^\pm)=\gamma^0\psi(0^\mp),
\]
the matching condition
$\psi^P(0^+)=T_\delta^P(q,\lambda)\psi^P(0^-)$
is automatically satisfied:
\begin{equation}
T_\delta^P(q,\lambda)\equiv \gamma^0T_\delta(q,\lambda)^{-1}\gamma^0=T_\delta(q,\lambda). \label{gen-delta-defP}
\end{equation}
Thus \eqref{gen-delta-defP} guarantees that $T_\delta(q,\lambda)$ remains invariant under parity such that the general fermionic $\delta$-potential is parity invariant, as it happens in the bosonic case.

\item Denoting $\psi^T(x,t)\equiv{\cal T}\psi(x,t)$, the matching condition \eqref{gen-delta-def} imposes over $\psi^T$ the time-reversal transformed matching conditon
\begin{equation}
\psi^T(0^+)=T_\delta^T(q,\lambda)\psi^P(0^-); \quad T_\delta^T(q,\lambda)=\gamma^0T_\delta(q,\lambda)^*\gamma^0.
\end{equation}
One immediately realizes that $T_\delta^T(q,\lambda)=T_\delta(q,\lambda)$. Hence the fermionic $\delta$-potential maintains the time-reversal invariance as in the scalar case.

\item The charge conjugated spinors $\psi^C\equiv {\cal C}\psi$ must satisfy the conjugated matching conditions:
\begin{equation}
\psi^C(0^+)=T_\delta^C(q,\lambda)\psi^C(0^-),
\end{equation}
where the charge-conjugated matching matrix is:
\begin{equation}
T_\delta^C(q,\lambda)=\gamma^2T_\delta(q,\lambda)^*\gamma^2=T_\delta(-q,\lambda).
\end{equation}
Thus, the fermionic $\delta$-potential as defined in \eqref{gen-delta-def} is not invariant under charge conjugation changing $q$ by $-q$ as long as $q\neq 0$.
%This result is what one expects considering that the coupling $q$ of the general fermionic $\delta$-potential represents an electric field as explained in Ref. \cite{sundberg}.
This result is what one expects implicitly considering the term with the coupling $q$ in \eqref{gen-delta-def} as an electrostatic potential~\cite{sundberg}.

\end{itemize}
 Therefore we are led to solve simultaneously the spectral problems for the Dirac Hamiltonian and its conjugate together with the $\delta$-matching condition and its conjugate:
\begin{eqnarray}
\boldsymbol{H}_D^{(0)}\psi(x)=\omega\psi(x);&&\psi(0^+)=T_\delta(q,\lambda)\psi(0^-),\label{ddse}\\
\overline{\boldsymbol{H}}_D^{(0)}\phi(x)=\omega\phi(x);&&\phi(0^+)=T_\delta(-q,\lambda)\phi(0^-) .\label{ddsp}
\end{eqnarray}
The eigenspinors of $\boldsymbol{H}_D^{(0)}$ with $\omega>m$ obeying the matching condition in \eqref{ddse}
correspond to electron scattering states, whereas those with $-m<\omega<m$ refer to electron bound states. On the contrary, the  $\overline{\boldsymbol{H}}_D^{(0)}$ eigenspinors  with $\omega>m$  complying with the matching condition in \eqref{ddsp} can be treated as positron scattering states,  but those with $-m<\omega<m$  we attribute to positron bound states .

The main objective of the present work is the study of this spectral problem in $1D$ relativistic quantum mechanics in order to build a fermionic quantum field theory system where the one-particle/antiparticle states are the eigenstates of $\boldsymbol{H}_D$/$\overline{\boldsymbol{H}}_D^{(0)}$.
The fermionic Fock space is thus constructed from these eigen-states instead of plane waves. In the next Section we introduce the necessary notation and basic formulas to understand the behaviour of fermions in a flat background without any external potential. In Section 3 we consider the dynamics of a relativistic 1D Fermi particle and antiparticle in one electrostatic $\delta$-potential  $V=q \delta(x) \boldsymbol{\mathbbm{1}}$ describing the effect of one impurity on the free propagation. Fermions (antifermions) are either trapped in bound states or distorted in scattering waves of $\boldsymbol{H}_D$ ($\overline{\boldsymbol{H}}_D$). The charge density of the bound states is also computed. In Section 4, the same study is performed for a mass-spike delta potential $V=\lambda \delta(x) \sigma_3$. A summary and outlook are  offered in the last section.

\section{Electron/positron propagation on a line}
We consider the one-dimensional Dirac field
\[
\Psi(t,x)=\left(\begin{array}{c} \psi_1(t,x) \\ \psi_2(t,x)\end{array}\right) \quad ; \quad \quad \psi_1(t,x):\, \mathbb{R}^{1,1}\, \rightarrow  \, \mathbb{C} \,  \, \, , \, \, \, \psi_2(t,x):\, \mathbb{R}^{1,1}\, \rightarrow  \, \mathbb{C},
\]
and set the following Dirac action:
\begin{eqnarray}
&&S_D=\int \, dt\int\, dx\left\{\bar{\Psi}(t,x)\left(i \gamma^\mu \partial_\mu-m \right)\Psi(t,x)\right\},%,\\
%&& S_D=\int \, dt\int\, dx\left\{\bar{\Psi}(t,x)\left(i \sigma_3 \partial_t-\sigma_2 \partial_x-m \right)\Psi(t,x)\right\}.
\label{daction}
\end{eqnarray}
in the Dirac representation of the Clifford algebra taking into account our choice of $\gamma$ matrices.
In the natural system of units, the dimension of the Dirac field $\Psi$ is: $[\Psi]=L^{-1/2}$. The Euler-Lagrange equation derived from the action (\ref{daction}) is the Dirac equation
\begin{eqnarray}
&&\hspace{2cm} i  \sigma_3\partial_t \Psi=\sigma_2 \partial_x \Psi +m\Psi, \label{dequ} \\ && \left(\begin{array}{cc} i\partial_t -m  & i\partial_x\\ -i \partial_x & -i \partial_t-m  \end{array}\right)\left(\begin{array}{c}\psi_1(t,x) \\ \psi_2(t,x)\end{array}\right)=\left(\begin{array}{c} 0 \\ 0\end{array}\right).\nonumber
\end{eqnarray}
The time-energy Fourier transform
\[
\psi_1(t,x)=\int\, d \omega \, e^{-i\omega t}\, \psi_1^\omega(x), \quad \quad \psi_2(t,x)=\int\, d\omega \, e^{-i\omega t}\, \psi_2^\omega(x),
\]
reduces the PDE Dirac equation to the ODE system
\begin{equation}
(\omega -m )\psi_1^\omega(x)+i\frac{d\psi_2^\omega}{dx}=0 \quad , \quad i\frac{d\psi_1^\omega}{dx}+(\omega+m)\psi_2^\omega(x)=0. \label{eq7}
\end{equation}
It is clear that the system (\ref{eq7}) is no more than the spectral equation for the quantum mechanical free Dirac Hamiltonian:
\begin{equation*}
\boldsymbol{H}_0  \Psi^\omega(x)=\omega \Psi^\omega(x) \quad, \qquad  \boldsymbol{H}_0=-i\sigma_1\frac{d}{dx}+m  \sigma_3 = \left(\begin{array}{cc} m  & -i \frac{d}{dx} \\ -i  \frac{d}{dx} & -m \end{array}\right),\label{rqmham}
\end{equation*}
acting on time-independent spinors{\footnote{We shall refer as spinor fields to the Fermi fields even though in one-dimension there is no spin.}. We remark now that a similar strategy based on time-energy Fourier transform also works when the effect of an external static potential, like those mentioned in the Introduction, is included in the action. The only required modification is to replace $\boldsymbol{H}_0$ by $\boldsymbol{H}_D$.

The analysis of free propagation also admits a position-momentum Fourier transform:
\[
\psi_1^\omega(x)=\int\, dk\, A(k) e^{ikx}, \quad  \quad \psi_2^\omega(x)=\int\, dk\, B(k) e^{ikx},
\]
which converts the ODE system (\ref{eq7}) in the algebraic homogeneous system
\begin{equation}
(\omega -m )\, A(k)-k \, B(k)=0, \quad  \quad  k \, A(k)-(\omega+m )\, B(k)=0. \label{seceq}
\end{equation}
%Observing the positive and negative energy eigenspinors when the fermion remains at rest
%\[
%\Psi^{+}(t,x)=A\left(\begin{array}{c} e^{-i mt}\\ 0\end{array}\right), \quad  \quad \Psi^{-}(t,x)=B\left(\begin{array}{c} 0 \\ e^{+ i mt}\end{array}\right) \, \, ,
%\]
%which solve (\ref{seceq}) if $k=0$, is easy to identify the non-trivial solutions of (\ref{seceq}).

Introducing the positive and negative energy eigenspinors which satisfy (\ref{seceq}) with $k=0$,
\[
\Psi^{+}(t,x)=A\left(\begin{array}{c} e^{-i mt}\\ 0\end{array}\right), \quad  \quad \Psi^{-}(t,x)=B\left(\begin{array}{c} 0 \\ e^{+ i mt}\end{array}\right) \, \, ,
\]
it is easy to derive the non-trivial solutions of (\ref{seceq}).

They occur if the following spectral condition holds
\begin{equation*}
{\rm det}\left(\begin{array}{cc} \omega -m  & -k \\ k & -(\omega + m )\end{array}\right) =0 \, \, \equiv \, \, \omega=\omega_\pm=\pm\sqrt{k^2  +m^2},
\end{equation*}
and the eigenspinors of moving electrons are classified in two classes:

(1) Positive energy $\omega_+$ electron spinor plane waves moving along the real axis with momentum $k\in \mathbb{R}$.
The solution of (\ref{seceq}), $B(k)=k/(\omega_+ + m )A(k)$, implies that the positive energy eigenspinors are
\begin{equation}\label{eq10}
\Psi^{+}(t,x;k)\!\! = \!\! A \, e^{-i \omega_+ t} e^{i k x} \, u_{+}(k)\, \, , \, \, \,
 u_{+}(k)= \left(\begin{array}{c} 1 \\ \frac{k}{\omega_+ + m } \end{array}\right).
\end{equation}

(2) Negative energy $\omega_-$ electron spinor plane waves moving along the real axis with momentum $k\in \mathbb{R}$.
We choose  the solution $A(k)=k/(\omega _- - m )B(k)$ of (\ref{seceq}) to find the negative energy eigenspinors
\begin{equation}\label{eq11}
\Psi^{-}(t,x;k)\!\! = \!\! B \, e^{-i \omega_- t}e^{ikx} u_-(k) \, \, , \, \, \,
u_-(k)=\left(\begin{array}{c} \frac{k}{\omega_- - m }\\ 1\end{array}\right).
\end{equation}
%The deep Dirac intelligence of negative energy electron states as positive energy positron states via the
%subtle concept of holes in the Dirac sea is implemented in this 1D context by replacing $u_-(k)$ by the
%positron spinors

In 1D space  the concept of the holes in the Dirac sea is implemented by replacing the negative energy spinors $u_-(k)$ with the positron spinors,
\begin{equation}\label{eq12}
v_+(k)=\gamma^2 u_+^*(k)=\left(\begin{array}{c} \frac{k}{\omega_++m} \\ 1\end{array}\right).
\end{equation}
which are solutions of the conjugate Dirac equation:
\begin{equation}
(\omega +m )\phi_1^\omega(x)+i\frac{d\phi_2^\omega}{dx}=0 \quad , \quad i\frac{d\phi_1^\omega}{dx}+(\omega-m)\phi_2^\omega(x)=0 \, \, . \label{eq71}
\end{equation}
Note that the $v_+(k)$ spinors are also orthogonal to the positive energy spinors $u_+(k)$. We thus describe the propagation of 1D fermions in terms of electron and positron plane waves:
\begin{eqnarray*}
\psi^+_{k}&\propto & u_+(k) e^{ikx} e^{-i\omega_+ t} \qquad  \qquad  \textrm{electron with momentum $k$, energy $\omega_+$}\\
\phi^+_{k}&\propto & v_+(k) e^{ikx} e^{-i\omega_+ t} \qquad \qquad \textrm{positron with momentum $k$, energy $\omega_+$}
\end{eqnarray*}
Therefore, from now on, we will work with the bihamiltonian system given in (\ref{ddse}-\ref{ddsp}), where the positron energy and the electron energy are always chosen as $\omega=+\sqrt{k^2+m^2}$.

\section{\lq\lq Electrostatic\rq\rq point delta-interaction}
Consider now a relativistic 1D fermion whose free propagation is disturbed by one impurity concentrated in one point that we describe by including a $\delta$-potential. The one-dimensional Dirac Hamiltonian with a single Dirac $\delta$-potential of \lq\lq electrostatic\rq\rq type is:
\begin{eqnarray*}
&&\hspace{-0.7cm} H_{\rm E\delta}=-i \sigma_1 \deruno{}{x} \,+ \,  m \sigma_3 +q  \delta(x) \mathbbm{1}, \quad   q=\nu \frac{e^2}{m^2}\in{\mathbb S}^1 \, , \, \, \,  \nu \in(0, 2\pi \frac{m^2}{e^2} ).
\end{eqnarray*}
Recall that $q$ is dimensionless: $[q]=1$. The spectral equation for this Hamiltonian $H_{\rm E\delta}\Psi(x)= \omega \Psi(x)$ is equivalent to the Dirac system of two first-order ODE's:
\begin{eqnarray}
-i \frac{d\psi_2}{dx} &=& (\omega -m)\psi_1(x),\label{deltadir1} \\ -i \frac{d\psi_1}{dx} &=& (\omega +m)\psi_2(x), \label{deltadir2}
\end{eqnarray}
%where the eigenspinors for the free Dirac Hamiltonian in zone I ($x<0$) and those in zone II ($x>0$), must be related across the %singularity implementing the \lq\lq electrostatic\rq\rq matching conditions at $x=0$  defined in (\ref{elec-d}):
where the eigenspinors for the free Dirac Hamiltonian in zone I ($x<0$) and those in zone II ($x>0$), must be related across the singularity at $x=0$ by the \lq\lq electrostatic\rq\rq matching conditions  defined in (\ref{elec-d}):
\begin{equation}
 \left(\begin{array}{c} \psi_1(0^+)\\ \psi_2(0^+)\end{array}\right) =\left(\begin{array}{rr} \cos q & -i\sin q\\ -i \sin q & \cos q\end{array}\right) \left(\begin{array}{c} \psi_1(0^-)\\ \psi_2(0^-)\end{array}\right).  \label{matchel1}
\end{equation}

%Simili modo,
Similarly, positron propagation disturbed by impurities that can be studied through $ \overline{H}_{\rm E\delta}\Phi(x)= \omega \Phi(x)$ is equivalent to the Dirac system of two first-order ODE's:
\begin{eqnarray}
-i \frac{d\phi_2}{dx} &=& (\omega +m)\phi_1(x),\label{deltadir3} \\ -i \frac{d\phi_1}{dx} &=& (\omega -m)\phi_2(x). \label{deltadir4}
\end{eqnarray}
The solution of the system \eqref{deltadir3}-\eqref{deltadir4} is identical to the solution of the previous system but the \lq\lq electrostatic\rq\rq matching conditions must be conjugated:
\begin{equation}
 \left(\begin{array}{c} \phi_1(0^+)\\ \phi_2(0^+)\end{array}\right) =\left(\begin{array}{rr} \cos q & i\sin q\\ i \sin q & \cos q\end{array}\right) \left(\begin{array}{c} \phi_1(0^-)\\ \phi_2(0^-)\end{array}\right).  \label{matchel2}
\end{equation}

Our goal is to search for,  bound states, i.e., $\vert \omega \vert< \vert m\vert$, and  scattering states, i.e., $\vert \omega \vert> \vert m\vert$,  both  for the case of electrons and positrons.

\subsection{Relativistic electron and positron bound states}
In order to compute bound states, exponentially decaying solutions of \eqref{deltadir1}-\eqref{deltadir2} system in zone I must be related to exponentially decaying solutions of the same system in zone II by implementing the electrostatic matching conditions \eqref{matchel1} to identify the electron bound states and identical procedure will provide positron bound states replacing the ODE system by \eqref{deltadir3}-\eqref{deltadir4} and using the matching condition \eqref{matchel2}.
\begin{itemize}

\item {\bf Zone I}: $x <0$
\begin{equation*}
\psi_1^{I}(x,\kappa)=A^{I}(\kappa)e^{\kappa x} \quad , \quad \psi_2^{I}(x,\kappa)=B^{I}(\kappa)e^{\kappa x} \, \, , \, \, \, \quad \kappa > 0.
\end{equation*}
Plugging this ansatz in the spectral equation system (\ref{deltadir1})-(\ref{deltadir2}) one finds a linear algebraic homogeneous system in $A^{I}$ and $B^{I}$ whose solution (taking into account that the value of the energy $\omega=\sqrt{m^2-\kappa^2}$ is compatible with bound states in zone I provided that $0<\kappa<\vert m\vert$) is the following  eigenspinor:
\begin{equation}
\Psi^{I}_+(x,\kappa)=A^{I}_+(\kappa)\left(\begin{array}{c} 1 \\ \frac{-i \kappa}{\omega + m}\end{array}\right) e^{\kappa x}. \label{freespinI}
\end{equation}

\item {\bf Zone II}: $x >0$
\begin{equation*}
\psi_1^{II}(x,\kappa)=A^{II}(\kappa)e^{-\kappa x} \quad , \quad \psi_2^{II}(x,\kappa)=B^{II}(\kappa)e^{-\kappa x} \, \, , \, \, \, \quad \kappa > 0.
\end{equation*}
Similarly, plugging this ansatz in the spectral equation system (\ref{deltadir1})-(\ref{deltadir2}) one finds a linear algebraic homogeneous system in $A^{II}$ and $B^{II}$ whose solution (taking into account the possible value of the energy $\omega$ compatible with the existence of bound states in zone II provided that $0<\kappa<\vert m\vert$) is the following  eigenspinor:
\begin{equation}
\Psi^{II}_+(x,\kappa)=A^{II}_+(\kappa)\left(\begin{array}{c} 1 \\ \frac{i \kappa}{\omega + m}\end{array}\right) e^{-\kappa x}.\label{freespinII}
\end{equation}
\end{itemize}

In order to join the eigenspinors in zone I with those in zone II at $x=0$, we impose  the  matching conditions at the origin (\ref{matchel1}) and obtain the linear homogeneous system:
\begin{equation}
\left(\begin{array}{cc}-\cos q+\frac{\kappa \sin q}{\omega + m} & 1 \\ i(\sin q+\frac{\kappa \cos q}{\omega+m}) &\frac{i\kappa}{\omega + m}\end{array}\right)\cdot \left(\begin{array}{c} A^{I}_+ \\ A^{II}_+\end{array}\right)=\left(\begin{array}{c} 0 \\0 \end{array}\right). \label{elelbssc}
\end{equation}
Non null solutions of the homogeneous system (\ref{elelbssc}) correspond to the roots of the determinant of the previous $2\times 2$ matrix. In the same way, imposing similar anstaz for the $\phi$ field and solving the spectral equation system  (\ref{deltadir3})-(\ref{deltadir4})  the possible eigenstates take the form
\begin{itemize}

\item {\bf Zone I}: $x <0$

\begin{equation}
\Phi^{I}_+(x,\kappa)=D^{I}_+(\kappa)\left(\begin{array}{c} \frac{-i \kappa}{\omega + m}\\ 1\end{array}\right) e^{\kappa x}. \label{freeposspin1}
\end{equation}

\item {\bf Zone II}: $x >0$
\begin{equation}
\Phi^{II}_+(x,\kappa)=D^{II}_+(\kappa)\left(\begin{array}{c} \frac{i \kappa}{\omega + m} \\ 1 \end{array}\right) e^{-\kappa x}. \label{freeposspin2}
\end{equation}
\end{itemize}
Relating the spinors in both zones through the matching condition for positrons in $x = 0$ (\ref{matchel2}) we obtain the linear homogeneous system
\begin{equation}
\left(\begin{array}{cc}-\cos q-\frac{\kappa \sin q}{\omega + m} & 1 \\ i(-\sin q+\frac{\kappa \cos q}{\omega+m}) &\frac{i\kappa}{\omega + m}\end{array}\right)\cdot \left(\begin{array}{c} D^{I}_+ \\ D^{II}_+\end{array}\right)=\left(\begin{array}{c} 0 \\0 \end{array}\right), \label{elelbssc1}
\end{equation}
that allow us to obtain the bound states of positrons.

Since the parameter $q$ is an angle and because the $\delta$-potential is defined by means of trigonometric functions, the signs of $\kappa, \omega$ change in every quadrant. The outcome is that there exists one bound state in each quadrant, two for electrons and two for positrons, distributed as follows
\begin{figure}[H]
\centering
\includegraphics[scale=0.5]{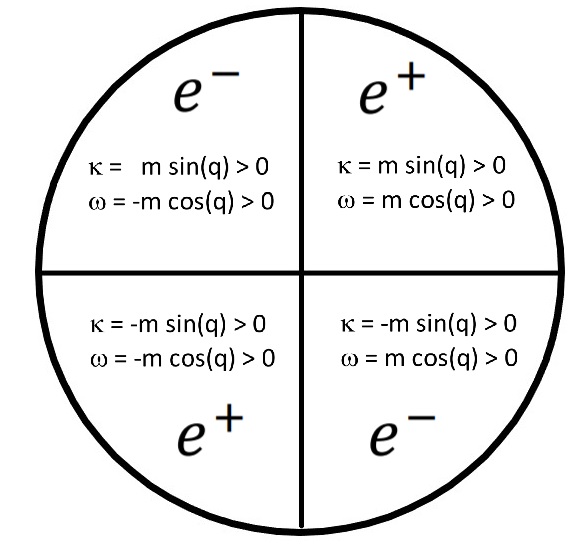}
\caption{\small Bound states for electrons and positrons in an ``electric" $\delta$-potential ($\lambda=0$).}
\end{figure}

These bound states correspond to one positive energy electron trapped in either one positive or negative energy state and one positive energy positron trapped alternatively in positive or negative energy states.  This structure is periodic in $q$.

\vspace{0.5cm}
\noindent {\bf Positron bound states}
\begin{enumerate}
\item $0<q<\frac{\pi}{2}$\, ; \, \, \, $\kappa_b= m \sin q > 0$ \, \, \, , \, \, \, $\omega_b=\sqrt{m^2-\kappa_b^2}= m \cos q > 0$
\begin{eqnarray}
D^{II}_+&=& D_+^{I} ,\nonumber \\
\Phi(x,\kappa_b)&=& D_+\,\left(\begin{array}{c} i \, {\rm sign}(x)\frac{ \sin q}{1+\cos q}\\ 1\end{array}\right) e^{-m\vert x\vert \sin q} .\label{eq34}
\end{eqnarray}

\item $\pi <q<\frac{3\pi}{2}$\, ; \, \, \, $\kappa_b= -m \sin q > 0$ \, \, \, , \, \, \, $\omega_b=\sqrt{m^2-\kappa_b^2}= -m \cos q > 0$
\begin{eqnarray}
D^{II}_+&=& -D_+^{I} ,\nonumber \\
\Phi(x,\kappa_b)&=& D_+\,\left(\begin{array}{c}  \frac{ i \sin q}{1-\cos q}\\ -{\rm sign}(x) \end{array}\right) e^{m\vert x\vert \sin q}. \label{eq34b}
\end{eqnarray}

\end{enumerate}

\noindent {\bf Electron bound states}
\begin{enumerate}
\item $\frac{\pi}{2}<q< \pi$\, ; \, \, \, $\kappa_b= m \sin q >0$ \,  , \, \, \, $\omega_b=\sqrt{m^2-\kappa_b^2}=  -m \cos q >0 $
\begin{eqnarray}
A^{II}_+&=&- A^{I}_+, \nonumber\\ \Psi(x,\kappa_b)&=& A_+ \left(\begin{array}{c} -{\rm sign}(x)  \\   \frac{ -i \, \sin q}{1-\cos q}\end{array}\right) e^{-m\vert x\vert \sin q}.\label{eq35a}
\end{eqnarray}

\item $\frac{3\pi}{2}<q<2 \pi$\, ; \, \, \, $\kappa_b= -m \sin q>0$ \,  , \, \, \, $\omega_b=\sqrt{m^2-\kappa_b^2}= m \cos q >0$
\begin{eqnarray}
A^{II}_+&=&A^{I}_+, \nonumber\\ \Psi(x,\kappa_b)&=& A_+ \left(\begin{array}{c}1 \\ -i\,  {\rm sign}(x)\frac{ \sin q}{1+\cos q}\end{array}\right) e^{m\vert x\vert \sin q}.\label{eq35}
\end{eqnarray}
\end{enumerate}

It is worthwhile to mention that if $q=\frac{\pi}{2}$ or $q=\frac{3\pi}{2}$ zero modes exist. For instance when $q=\frac{\pi}{2}$ we have $\kappa_b=m$,
$\omega_b=0$ whereas the eigenspinor reads:
\[
\Phi(x,m)=D_+\left(\begin{array}{c} i\, \textrm{sign} (x) \\ 1 \end{array}\right) e^{-m\vert x\vert} \, \, .
\]
We stress that the bound states just described are closer to the bound states in the scalar case with mixed potential
of the form $V(x)=-a \delta(x)+b\delta^\prime(x)$, see \cite{GNN}. In both cases the normalizable wave functions exhibit finite discontinuities at the origin.

\subsection{On the charge density}

The charge density can be written as:
\begin{eqnarray}\label{dens}
 j^0(t,x)&=& \pm Q \, \overline{\varphi}(t,x)\gamma^0 \varphi(t,x) = \pm Q \, \varphi^\dagger(t,x)\varphi(t,x) \nonumber\\&=&  \pm Q \, \left(\varphi_1^*(t,x)\varphi_1(t,x)+\varphi_2^*(t,x)\varphi_2(t,x)\right),
\end{eqnarray}
being $Q$ a positive constant and taken into account that $+$ will be chosen in the case of electrons and ${}-{}
$ for positrons. On the one hand, if we substitute the positron bound states (\ref{eq34}, \ref{eq34b}) in (\ref{dens}) the charge density obtained  is
\begin{eqnarray}
&&j_0(x)= -m \, Q\, \sin q\,  e^{-2 m |x| \sin q},  \qquad \qquad \textrm{iff} \qquad 0<q<\frac{\pi}{2}. \label{dens1}\\
&&j_0(x)= +m \, Q\, \sin q\,  e^{2 m |x| \sin q},  \qquad \qquad \,\,\, \textrm{iff} \qquad \pi<q<\frac{3\pi}{2}.\label{dens1b}
\end{eqnarray}
%Both results are shown in Figure 2,4 respectively.
On the other hand, if we substitute the electron bound states  (\ref{eq35a}, \ref{eq35}) in (\ref{dens}) the charge density obtained  is
\begin{eqnarray}
&&j_0(x)=+m \, Q\, \sin q\,  e^{-2 m |x| \sin q},  \qquad \qquad \textrm{iff} \qquad \frac{\pi}{2}<q<\pi.\label{dens2b}\\
&&j_0(x)=-m \, Q\, \sin q\,  e^{2 m |x| \sin q},  \qquad \qquad  \,\,\, \textrm{iff} \qquad \frac{3\pi}{2}<q<2\pi.\label{dens2}
\end{eqnarray}
All the results are shown in Figure  2.

\begin{figure}[H]
\centering
\includegraphics[scale=0.3]{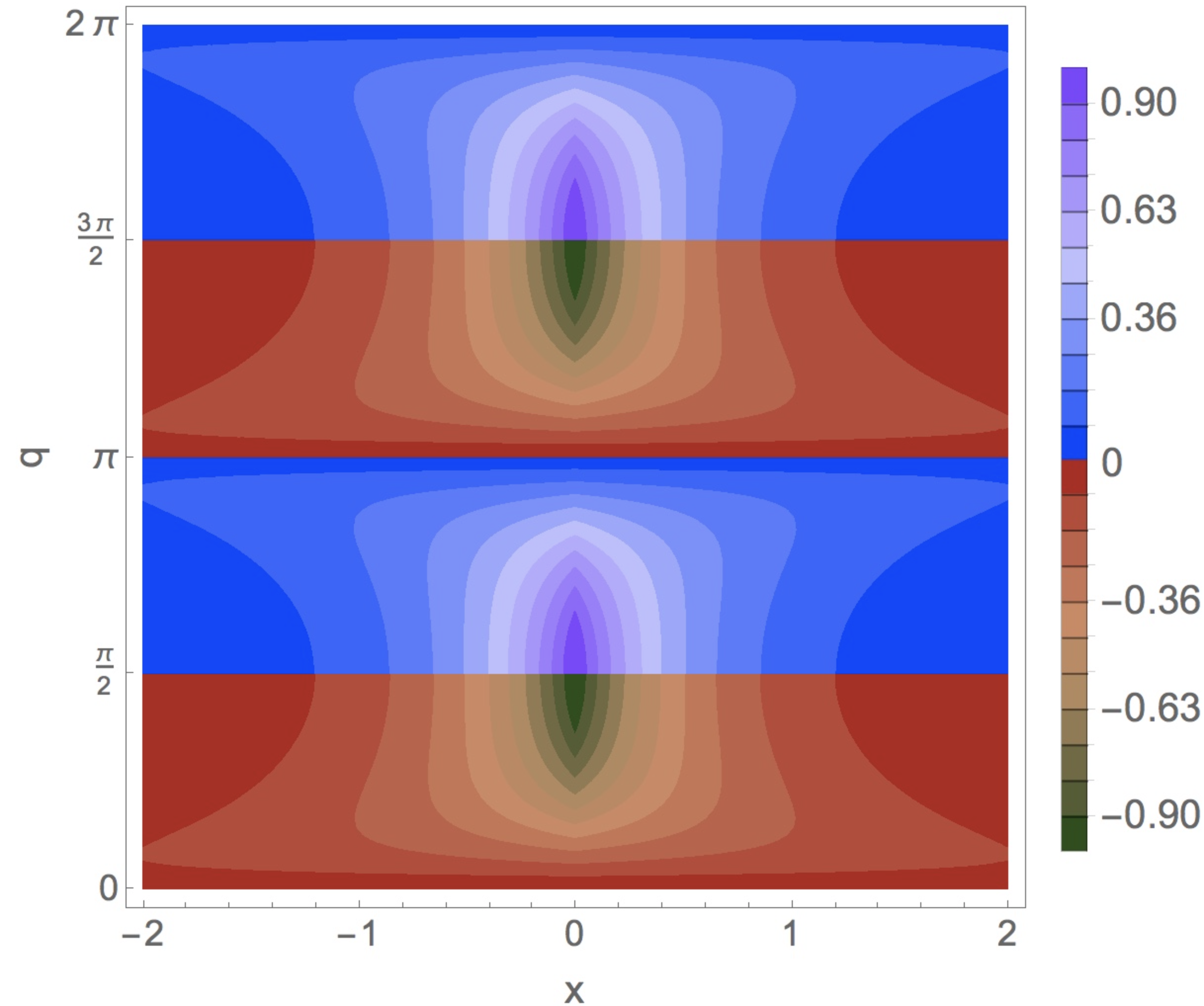}
\caption{\small Charge density as a function of $x$ when $Q=m=1$ and $\lambda =0$ (``electric" $\delta$-potential). For $0<q<\pi/2$ and  $\pi<q<3\pi/2$ the charge densities of a positron bound state are plotted (\eqref{dens1}, \eqref{dens1b} respectively). For $\pi/2<q<\pi$ and  $3\pi/2<q<2\pi$ the charge densities of a electron bound state are plotted (\eqref{dens2b}, \eqref{dens2} respectively).}
\end{figure}

%\begin{figure}[H]
%\centering
%\includegraphics[scale=0.6]{de2}
%\caption{\small Charge density for electrons (\ref{dens2b}) as a function of $x$ when $Q=m=1$ and $\lambda =0$ (``electric" $\delta$-potential).}
%\end{figure}

%\begin{figure}[H]
%\centering
%\includegraphics[scale=0.6]{dp3}
%\caption{\small Charge density for positrons (\ref{dens1b})  as a function of $x$ when $Q=m=1$ and $\lambda =0$ (``electric" $\delta$-potential).}
%\end{figure}

%\begin{figure}[H]
%\centering
%\includegraphics[scale=0.6]{de4}
%\caption{\small Charge density for electrons (\ref{dens2})  as a function of $x$ when $Q=m=1$ and $\lambda =0$ (``electric" $\delta$-potential).}
%\end{figure}

\subsection{Relativistic electron and positron scattering spinors}
We pass to study the scattering of 1D Dirac particles through a Dirac electrostatic $\delta$-potential ($V(x)=q \delta(x) \mathbbm{1}$) in order to obtain the scattering amplitudes. On the one hand, electron scattering spinors coming from the left towards the $\delta$-impurity   (\lq\lq diestro\rq\rq scattering) have the form:
\begin{equation}	\label{eq20}
\Psi^R (x,k) = \left\{\begin{array}{l} u_+(k)\, e^{ikx} + \rho_R(k) \, \gamma^0 u_+(k) \,  e^{-ikx}, \qquad \qquad \, \, \, \,  x <0\\ \sigma_R(k) \, u_+(k)\,  e^{ikx}, \qquad \qquad \qquad \qquad \qquad \, \, \, \, \quad x>0\end{array}\right. ,
\end{equation}
where $u_+(k)$ is the positive energy electron spinor that solves the free static Dirac equation for plane waves moving along the real line (\ref{eq10}). The solutions in both zones are related at the origin through the electrostatic $\delta$-matching conditions (\ref{elec-d}) if and only if the transmission and reflection scattering amplitudes are:
\begin{equation}\label{eq41}
\sigma_R(k)= \frac{k}{k\cos q + i\sqrt{k^2+m^2} \sin q},
\quad  \quad \rho_R(k)= -\frac{i \, \,m \sin q}{k\cos q +i \sqrt{k^2+m^2} \sin q},
\end{equation}
which obviusly satisfy the unitarity condition: $\Big\vert \sigma_R(k)\Big\vert^2+\Big\vert \rho_R(k)\Big\vert^2=1$.
On the other hand, electrons coming from the right towards the $\delta$-impurity, (\lq\lq zurdo\rq\rq scattering), are described by spinors of the form
\begin{equation}
\Psi^L (x,k) = \left\{\begin{array}{l} \sigma_L(k) \, \gamma^0 u_+(k) \,  e^{-ikx}, \qquad \qquad \qquad \quad \, \, \,  x<0\\ \rho_L(k) \, u_+(k)\,  e^{ikx} + \gamma^0 u_+(k) \, e^{-ikx},\qquad  \, x>0\end{array}\right. \label{scattrans},
\end{equation}
The $\delta$-well matching conditions  (\ref{elec-d}) for this  ``zurdo''  scattering ansatz are satisfied if $\sigma_R(k)=\sigma_L(k)$ and $\rho_R(k)=\rho_L(k)$.
It is worth noting that
\begin{itemize}
\item Since the scattering amplitudes for \lq\lq diestro\rq\rq and \lq\lq zurdo\rq\rq   scattering of the electrons through an electrostatic $\delta$-potential are identical, the processes governed by this potential are parity and time-reversal invariants.

\item Purely imaginary poles $k=i \kappa$ of the transmission amplitude $\sigma$ are the bound states of the spectrum
if $\kappa$ is real and positive. In formula \eqref{eq41} we observe that poles of this type appear if the imaginary momentum satisfies the equation
\[
\frac{\kappa_b}{\sqrt{m^2-\kappa_b^2}}=-\tan q,
\]
which admits positive solutions for $\kappa_b$ only if $\tan q <0$, i.e. if $q$ lives in the second or fourth quadrant.
Moreover, explicit solutions of the previous equation are: $\kappa_b=\pm m \sin q$, i.e. assuming that $m>0$ the plus sign must be selected in the second quadrant and the minus sign is valid in the fourth quadrant.

\item Probability is conserved even in this relativistic quantum mechanical context provided that $\omega^2>m^2$, $m>0$.
\end{itemize}

For positrons, the \lq\lq diestro\rq\rq scattering ansatz of the spinor (that is a solution of the conjugate Dirac equation in zones I and II) is
\begin{equation}\label{eq22}
\Phi^R (x,k) = \left\{\begin{array}{l} v_+(k)\, e^{-ikx} - \tilde{\rho}_R(k) \,\gamma^0 v_+(k) \,  e^{ikx}, \qquad \quad \, x<0 \\ \tilde{\sigma}_R(k) \,  v_+(k)\,  e^{-ikx}, \, \, \, \quad \qquad \qquad \qquad \qquad x>0\end{array}\right. ,
\end{equation}
where $v_+(k)$ is the positive energy positron spinor that represents plane  waves moving along the real line (\ref{eq12}).
By imposing the matching conditions on $x = 0$ (\ref{ddsp}),  the following scattering coefficients are obtained
\begin{equation}
\tilde{\sigma}_R(k)= \frac{k}{k\cos q - i\sqrt{k^2+m^2} \sin q},
\quad  \quad \tilde{\rho}_R(k)= \frac{i \, \,m \sin q}{k\cos q -i \sqrt{k^2+m^2} \sin q} \, \, .\label{eq23}
\end{equation}
Again, unitarity is preserved: $\Big\vert \tilde{\sigma}(k)\Big\vert^2+\Big\vert \tilde{\rho}_R(k)\Big\vert^2=1$.

The \lq\lq zurdo\rq\rq positron scattering ansatz, however, takes the form
\begin{equation}
\Phi^L (x,k) = \left\{\begin{array}{l} -\tilde{\sigma}_L(k) \,  \gamma^0 v_+(k) \,  e^{ikx}, \qquad \qquad \qquad \qquad \qquad x<0\\ \tilde{\rho}_L(k) \,  v_+(k)\,  e^{-ikx} -  \gamma^0 v_+(k) \, e^{ikx},\qquad \qquad \,  \, \, \,  \, x>0\end{array}\right. ,
\end{equation}
Again, by imposing the matching conditions on $x = 0$ (\ref{ddsp}),  we find that $\tilde{\sigma}_L(k)=\tilde{\sigma}_R(k)$ and $\tilde{\rho}_L(k)=\tilde{\rho}_R(k)$. It is worth noting that
\begin{itemize}
\item The scattering amplitudes for ``diestro'' and ``zurdo'' scattering of positrons through an electrostatic $\delta$-well are identical: there is no violation of parity and time reversal invariance in the
scattering of positrons by $\delta$-impurities.

\item The purely imaginary $k=i\kappa$ poles of $\tilde{\sigma}$ are the positron bound states in the spectrum of the conjugate Dirac Hamiltonian.
In formula (\ref{eq23}) we observe that poles of this type appear if the imaginary momentum satisfies the equation
\[
\frac{\kappa_b}{\sqrt{m^2-\kappa_b^2}}=\tan q,
\]
which admits positive solutions for $\kappa_b$ only if $\tan q >0$, i.e. if $q$ lives in the first or third quadrant. Between the  explicit solutions of this eqution, $\kappa_b=\pm m \sin q$, the plus sign must be chosen in the first quadrant, and the minus sign in the third quadrant assuming that $m>0$.

\item The relations between diestro and zurdo scattering amplitudes for electrons and positrons are as follows:
\begin{eqnarray}
&&\sigma_R(k)=\sigma_L(k)=\tilde{\sigma}^*_R =\tilde{\sigma}^*_L(k),\nonumber\\
&&\rho_R(k)=\rho_L(k)=\tilde{\rho}^*_R(k)=\tilde{\rho}^*_L(k).
\end{eqnarray}

\end{itemize}
The unitary $\boldsymbol{S}$-matrix
\[
\boldsymbol{S}=\left(\begin{array}{cc}\sigma(k) & \rho(k) \\ \rho(k) & \sigma(k)\end{array}\right) \quad , \quad \boldsymbol{S}^\dagger\cdot \boldsymbol{S}=\boldsymbol{I},
\]
encodes the phase shifts in its spectrum; $\lambda_\pm = \sigma \pm \rho=e^{2i\delta_\pm(k)}$. The phase shifts $\delta_\pm(k)$ in the even and odd channels are thus
\[
\tan 2\delta_\pm(k)=\frac{{\rm Im}(\sigma(k)\pm \rho(k))}{{\rm Re}(\sigma(k)\pm \rho(k))},
\]
whereas the total phase shift $\delta(k)=\delta_+(k)+\delta_-(k)$ is easily derived:
\begin{equation*}
\tan 2\delta(k) = \frac{\textrm{Im}[\sigma^2(k)-\rho^2(k)]}{\textrm{Re}[\sigma^2(k)-\rho^2(k)]}=\frac{2k \sqrt{k^2+m^2} \sin (2q)}{m^2-(2k^2+m^2)\cos(2q)}.
\end{equation*}

\section{\lq\lq Mass-spike\rq\rq $\delta$-potential}
Next, we consider the one-dimensional Dirac Hamiltonian with a single Dirac $\delta$-potential disturbing the
mass term:
\begin{eqnarray*}
&&\hspace{-0.7cm} H_{\rm M\delta}=-i  \sigma_1 \frac{d}{dx} \, + \,  (m  +\lambda \delta(x)) \sigma_3.
\end{eqnarray*}

\subsection{Relativistic bound states in mass-spike $\delta$ wells}
Firstly, we will search for bound states where electrons and positrons are trapped by mass-spike $\delta$ wells. In the case of electrons, away from the singularity the positive energy spinors take the form (\ref{freespinI}, \ref{freespinII}).  The continuation to the whole real line is achieved by applying to those spinors the relativistic matching conditions at the origin $x=0$ (\ref{eq9}) as follows:
\begin{eqnarray}\label{eq27}
&& \left(\begin{array}{c}\psi_{1+}^{II}(0,\kappa) \\ \psi_{2+}^{II}(0,\kappa)\end{array}\right)= \left(\begin{array}{cc} \cosh \lambda & i \, \sinh \lambda \\ -i \, \sinh \lambda & \cosh \lambda \end{array}\right) \left(\begin{array}{c}\psi_{1+}^{I}(0,\kappa) \\ \psi_{2+}^{I}(0,\kappa)\end{array}\right).
\end{eqnarray}
In this way, we obtain the following homogeneous algebraic system written in matrix form as:
\begin{equation}
\left(\begin{array}{cc}  -\left(\cosh \lambda + \frac{\kappa \sinh \lambda}{m+\sqrt{m^2-\kappa^2}}\right) & 1 \\ & \\ i \left(\frac{\kappa \cosh \lambda}{m+\sqrt{m^2-\kappa^2}}+\sinh \lambda\right) & \frac{i \kappa}{m+\sqrt{m^2-\kappa^2}}\end{array}\right)\left(\begin{array}{c}A_+^{I}\\ \\ A_+^{II}\end{array}\right)=\left(\begin{array}{c} 0 \\ \\ 0\end{array}\right). \label{spinbshom}
\end{equation}
The determinant of this $2\times 2$ matrix is zero such that there is a non null solution of \eqref{spinbshom} if $\kappa_b=- m\tanh\lambda$ which provides a normalizable spinor only if $\lambda<0$. The energy of the electron bound state is $\omega_b= m \, \textrm{sech}\lambda$. This bound state spinor  is extended to the whole line by means of the condition $A_+^{II}= A_+^{I}$, i.e. the spinor takes the form:
\begin{equation}\label{eq48}
\Psi_+(x,\kappa_b)=A_+ \left(\begin{array}{c}1 \\ -i \, {\rm sign} (x)\, \frac{\sinh \lambda}{1+\cosh \lambda }\end{array}\right) e^{m \vert x\vert\tanh  \lambda}.
\end{equation}
The charge density of this bound state is obtained by replacing the spinor (\ref{eq48}) in the equation (\ref{dens}), arriving at the result
\begin{equation}\label{dens3}
j_0(x)= -m \, Q \tanh \lambda \, e^{2m \vert x \vert \tanh \lambda},
\end{equation}
which is represented in Figure 3.
\begin{figure}[H]
\centering
\includegraphics[scale=0.35]{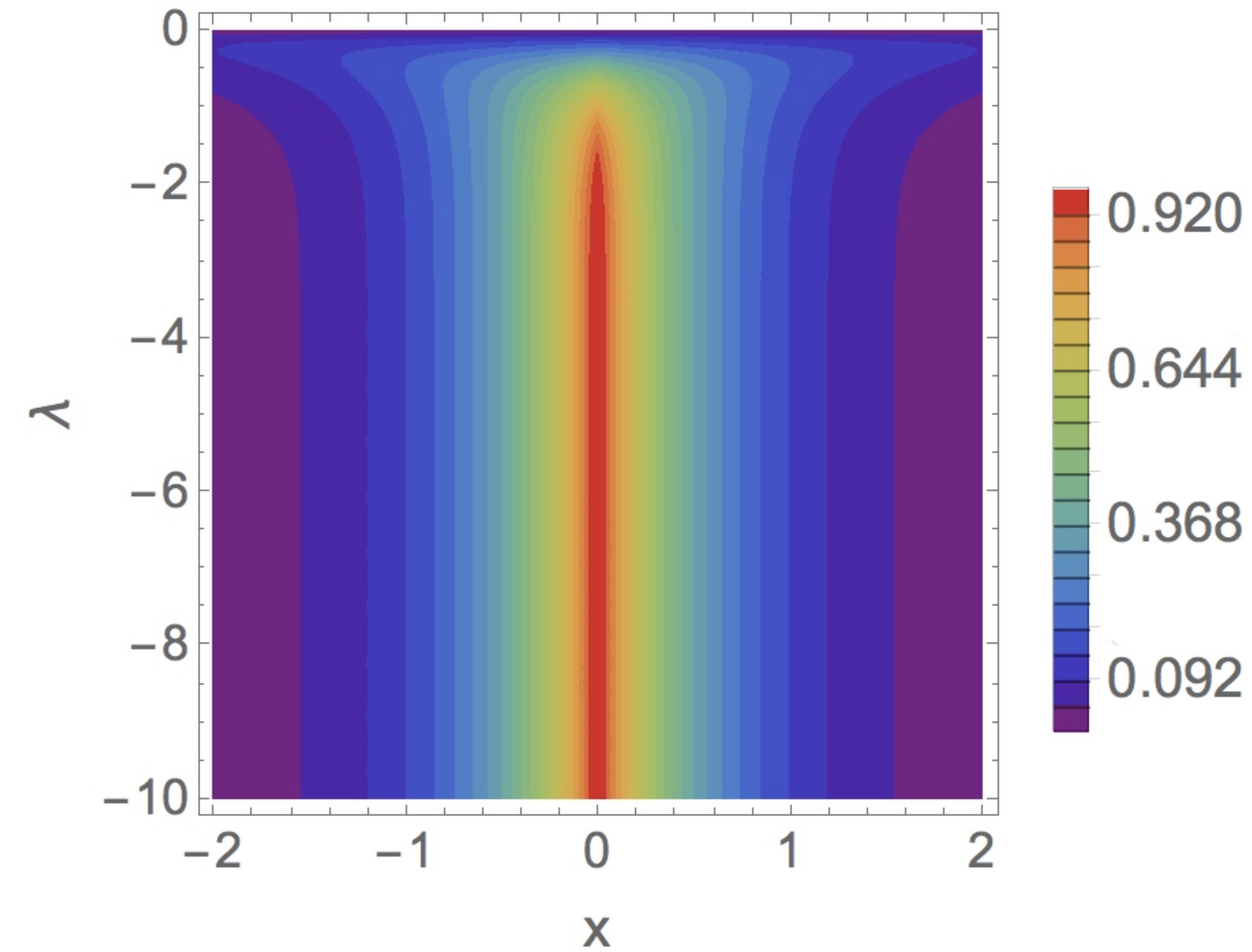}
\caption{\small Charge density (\ref{dens3}) as a function of $x$ for electrons when $Q=m=1$ and $q =0$ (``massive" $\delta$-potential).}
\end{figure}

Investigation of positron bound states requires applying the same relativistic matching conditions (\ref{eq9}) at the origin to the bound state positron spinors  \eqref{freeposspin1}-\eqref{freeposspin2} in order to obtain the following homogeneous algebraic system written in matrix form as:
\begin{equation}
\left(\begin{array}{cc} i \left(\frac{\kappa \cosh \lambda}{m+\sqrt{m^2-\kappa^2}}-\sinh \lambda\right) & \frac{i \kappa}{m+\sqrt{m^2-\kappa^2}} \\ &\\  -\cosh \lambda + \frac{\kappa \sinh \lambda}{m+\sqrt{m^2-\kappa^2}} & 1 \end{array}\right)\left(\begin{array}{c}D_+^{I}\\ \\ D_+^{II}\end{array}\right)=\left(\begin{array}{c} 0 \\ \\ 0\end{array}\right). \label{spinbshom1}
\end{equation}
The determinant of this $2\times 2$ matrix is zero such that there is a non null solution of \eqref{spinbshom1} if $\kappa_b=m\tanh\lambda$. This imaginary momentum provides a normalizable spinor only if $\lambda>0$. The energy of this positron bound state is $\omega_b=m\, \textrm{sech}\,  \lambda$. The spinor bound state is extended to the whole line by means of the  condition $D_+^{II}=D_+^{I}$:
\begin{equation}\label{eq51}
\hspace{-0.25cm}\Phi_+(x,\kappa_b)=D_+\left(\begin{array}{c} i \, {\rm sign} (x)\, \frac{\sinh \lambda}{1+\cosh \lambda }\\ 1\end{array}\right) e^{-m \vert x\vert\tanh \lambda }.
\end{equation}
The charge density of this bound state is obtained by replacing the spinor (\ref{eq51}) in the equation (\ref{dens}), arriving at the result
\begin{equation}\label{dens4}
j_0(x)= -m \, Q \tanh \lambda \, e^{-2m \vert x \vert \tanh \lambda},
\end{equation}
which is represented in Figure 4.
\begin{figure}[H]
\centering
\includegraphics[scale=0.35]{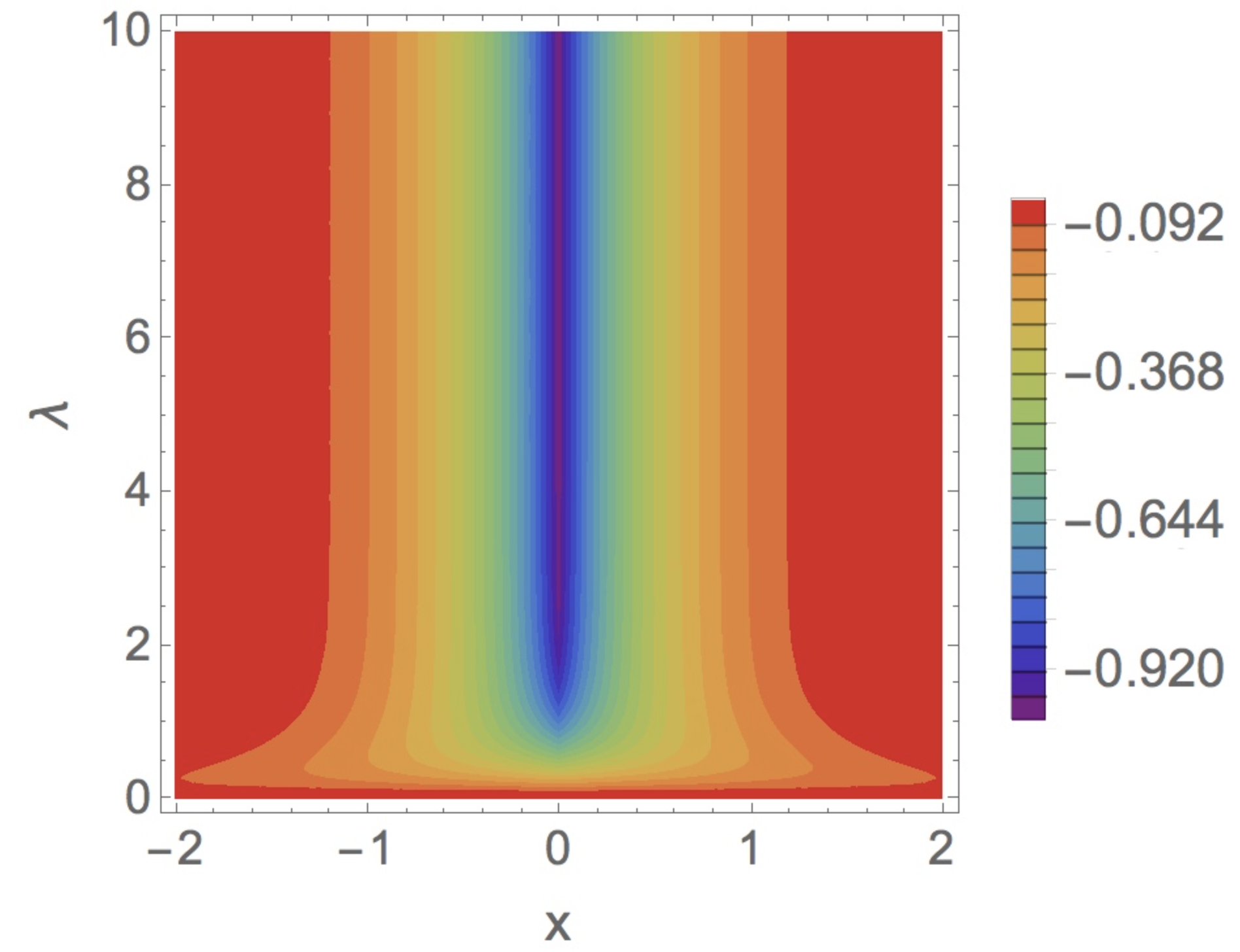}
\caption{\small Charge density (\ref{dens4}) as a function of $x$ for positrons when $Q=m=1$ and $q =0$ (``massive" $\delta$-potential).}
\end{figure}

\subsection{Electron and positron scattering spinors}
To obtain the scattering amplitudes for electrons coming from the left (\lq\lq diestro\rq\rq scattering) on mass-spike impurities the free spinors in zones I and II away from the origin (\ref{eq20}) must be joined by using the $\mathbb{SU}(1,1)$ matrix appearing in \eqref{eq27}. More explicitly, this matching sets an algebraic system whose solutions are the scattering coefficients:
\begin{equation*}
\sigma_R(k)= \frac{k}{k \cosh \lambda + i m\sinh \lambda}\, , \, \, \, \, \rho_R(k)= \frac{-i \sqrt{k^2+m^2}\sinh \lambda}{k \cosh \lambda + im\sinh \lambda},
\end{equation*}
which obviously respect probability conservation:
\[
\Big\vert \sigma_R(k)\Big\vert^2+\Big\vert \rho_R(k)\Big\vert^2=1.
\]
Repeating this procedure for the \lq\lq zurdo\rq\rq scattering spinorial ansatz (\ref{scattrans}) we conclude with the  same relativistic $\delta$-interaction scattering amplitudes as in \lq\lq diestro\rq\rq scattering. In sum
\begin{itemize}
\item The scattering amplitudes for diestro and zurdo scattering of electrons through a mass-spike $\delta$-interaction are identical. This means that the mass-spike $\delta$ interaction respect both parity and time-reversal symmetries.

\item The $\boldsymbol{S}$-matrix is unitary and the phase shifts appear as the exponents of its eigenvalues. The total phase shift is:
\begin{equation*}
\tanh 2\delta(k)=\frac{{\rm Im}[\sigma^2(k)-\rho^2(k)]}{{\rm Re}[\sigma^2(k)-\rho^2(k)]}=\frac{-2 k m \sinh 2\lambda}{k^2+m^2+ (k^2-m^2)\cosh 2\lambda}\, \, .
\end{equation*}

\item The purely imaginary poles of the transmission amplitude $\sigma(k)$ with positive imaginary part are the bound states of the spectrum and occur when: $k_b=i\kappa_b=-i m \tanh \lambda$, i.e., $\omega_b= m \, \textrm{sech} \lambda$. It must be fulfilled that $\tanh \lambda < 0$.

\end{itemize}
Investigation of the  positron \lq\lq diestro\rq\rq scattering amplitudes is achieved by imposing the relativistic matching  conditions
\begin{eqnarray*}
\Phi^{II}_+(0,k)&=&\left(\begin{array}{cc}\cosh \lambda & i \sinh \lambda \\ -i \sinh \lambda & \cosh \lambda\end{array}\right)\cdot \Phi_+^{I}(0,k),
\end{eqnarray*}
on the positron spinor scattering ansatz \eqref{eq22}. This criterion is tantamount to an algebraic system for the scattering amplitudes whose solutions are:
\begin{equation*}
\tilde{\sigma}_R(k)= \frac{k}{k \cosh \lambda -i m \sinh \lambda}\, , \, \, \, \, \tilde{\rho}_R(k)= \frac{i \sqrt{k^2+m^2} \sinh \lambda}{k \cosh \lambda -i m\sinh \lambda},
\end{equation*}
which also respects probability conservation:
\begin{equation*}
\Big\vert \tilde{\sigma}_R(k)\Big\vert^2+\Big\vert \tilde{\rho}_R(k)\Big\vert^2=1 \, \, .
\end{equation*}
To avoid repetitions, we skip a detailed computing of the positron scattering amplitudes for \lq\lq zurdo\rq\rq scattering, we merely states that are identical to the scattering amplitudes for positrons coming from the left towards the $\delta$-obstacle. Thus we summarize the main features of positron scattering through a mass-spike $\delta$-imputity as follows:

\begin{itemize}
\item The scattering amplitudes for diestro and zurdo scattering of positrons through a mass-spike $\delta$-interaction are identical to each other. Therefore, there is no breaking of either parity or time-reversal symmetries.

\item The $\boldsymbol{S}$-matrix is unitary and the phase shifts appears as the exponents of its eigenvalues. The total phase shifts for positrons are

\begin{equation*}
\tanh 2\tilde{\delta}(k)=\frac{{\rm Im}[\tilde{\sigma}^2(k)-\tilde{\rho}^2(k)]}{{\rm Re}[\tilde{\sigma}^2(k)-\tilde{\rho}^2(k)]}=\frac{2 k m\sinh 2\lambda}{ k^2+m^2+(k^2-m^2)\cosh 2\lambda}\, \, .
\end{equation*}

\item The purely imaginary poles of the transmission amplitude $\tilde{\sigma}(k)$ with positive imaginary part are the positron bound states of the spectrum and occur when $k_b=i\kappa_b= i m\tanh\lambda$, i.e., $\omega_b= m \, \textrm{sech}\lambda$. It must be fulfilled that $\tanh \lambda >0$.

\item The relations between diestro and zurdo scattering amplitudes for electrons and positrons in a mass-spike $\delta$-potential are as follows:
\begin{eqnarray}
&&\sigma_R(k)=\sigma_L(k)=\tilde{\sigma}^*_R =\tilde{\sigma}^*_L(k),\nonumber\\
&&\rho_R(k)=\rho_L(k)=\tilde{\rho}^*_R(k)=\tilde{\rho}^*_L(k).
\end{eqnarray}

\end{itemize}

\section{Summary and outlook}
The spectrum of the 1D Dirac Hamiltonian providing the one-particle spectrum for 1D electrons and positrons has been analyzed when there is one impurity that distorts the free propagation of fermions. We have implemented the impurity by means of Dirac $\delta$-potentials of two types that we denote respectively as electrostatic and mass-spike according to their physical meaning.

In the electrostatic case (where the coupling is an angle) we find that:
\begin{enumerate}

\item There are two quadrants ($\pi/2<q<\pi$ and $3\pi/2<q<2\pi$) where the coupling of the $\delta$-interaction gives rise to one electron bound state.
One positron bound state arise in the other two quadrants ($0<q<\pi/2$ and $\pi<q<3\pi/2 $).

\item Regarding scattering amplitudes we found that positrons and electrons are scattered by the impurity so that the electron scattering coefficients are the conjugate of the positron ones.
\end{enumerate}

For mass-spike $\delta$-potentials our results are:
\begin{enumerate}
\item There is one bound state of electrons if the coupling is negative and other one of positrons if the coupling is positive.

\item The scattering amplitudes of electrons due to a mass-spike $\delta$-impurity are the conjugate of positrons ones.

\end{enumerate}

 We plan to continue this investigation along the following lines of research:

 \begin{itemize}

 \item First, our purpose is to study the effect on free fermions of an impurity carrying both electrostatic and
 mass-spike couplings.

 \item Second, it is our intention to consider two, several or even infinity $\delta$-impurities (often called as $\delta$-comb potential), as the periodic potentials arise in various materials models.

 \item Third, after having managed all these tasks, we envisage to compute quantum vacuum energies and Casimir forces
 induced by these 1D fermions, in a parallel analysis to that performed for bosons in \cite{JMMCJMG} and References quoted therein.

 \end{itemize}

\section*{Acknowledgmentss}
JMG, JMMC and LSS are grateful to the Spanish Government-MINECO (MTM2014-57129-C2-1-P) and the Junta de Castilla y Le\'on (BU229P18, VA137G18 and VA057U16) for the financial support. LSS is grateful the Valladolid University, through the PhD fellowships programme ({\it ``Contratos predoctorales de la UVa''}). JMMC would like to thank M. Gadella, G. Fucci, M. Asorey and L. M. Nieto for fruitful discussions. JMMC and LSS acknowledge the technical computing support received by Diego Rogel.

\bibliography{bibliography-dirac_delta_diracf}{}
\bibliographystyle{unsrt}

\end{document}